# xSCYTE: Express Single-frame Cytometer through Tomographic Phase

Baoliang Ge[1,2,†], Yanping He[3,4,5,†], Mo Deng[1,†], Md Habibur Rahman[3], Yijin Wang[3], Ziling Wu[1], Yongliang Yang[6], Cuifang Kuang[7], Chung Hong N. Wong[8,9], Michael K. Chan[8,9], Yi-Ping Ho[3,9,10], Liting Duan[3], Zahid Yaqoob[2], Peter T. C. So[1,2,11,12,*], George Barbastathis[1,12,13,*], and Renjie Zhou[3,*]

[1]*Department of Mechanical Engineering, Massachusetts Institute of Technology, Cambridge, MA 02139, USA*

[2]*Laser Biomedical Research Center, Massachusetts Institute of Technology, Cambridge, MA 02139, USA*

[3]*Department of Biomedical Engineering, The Chinese University of Hong Kong, Shatin, New Territories, Hong Kong SAR, China*

[4]*School of Artificial Intelligence Science and Technology, University of Shanghai for Science and Technology, Shanghai, 200093, China*

[5]*Institute of Photonic Chips, University of Shanghai for Science and Technology, Shanghai, 200093, China*

[6]*The State Key Laboratory of Robotics, Shenyang Institute of Automation, Chinese Academy of Sciences, Shenyang 110016, China*

[7]*State Key Laboratory of Extreme Photonics and Instrumentation, College of Optical Science and Engineering, Zhejiang University, Hangzhou 310027, China*

[8]*School of Life Sciences, The Chinese University of Hong Kong, Shatin, New Territories, Hong Kong SAR, China*

[9]*Centre for Novel Biomaterials, The Chinese University of Hong Kong, Shatin, New Territories, Hong Kong SAR, China*

[10]*Hong Kong Branch of CAS Center for Excellence in Animal Evolution and Genetics, The Chinese University of Hong Kong, Shatin, New Territories, Hong Kong SAR, China*

[11]*Department of Biological Engineering, Massachusetts Institute of Technology, Cambridge, MA 02139, USA*

[12]*Interdisciplinary Research Group on Critical Analytics for Manufacturing Personalized Medicine (CAMP), Singapore-MIT Alliance for Research and Technology Centre, Singapore*

[13]*Intra-Create Thematic Grant on Retinal Analytics through Machine Learning aiding Physics (RAMP), Singapore-MIT Alliance for Research and Technology Centre, Singapore*

[†]*These authors contributed equally to this work*

**Corresponding authors: ptso@mit.edu; gbarb@mit.edu; rjzhou@cuhk.edu.hk*

**Abstract:** Rapid, comprehensive, and accurate cell phenotyping without compromising viability, is crucial to many important biomedical applications, including stem-cell therapy, drug screening, and liquid biopsy. Typical image cytometry methods acquire two-dimensional (2D) fluorescence images, where the fluorescence labelling process may damage living cells, and the information from 2D images is not comprehensive enough for precise cell analysis. Although three-dimensional (3D) label-free image cytometry holds great promise, its high throughput development faces several technical challenges. Here, we report eXpress Single-frame CYtometer through Tomographic phasE (xSCYTE), which reconstructs 3D Refractive Index (RI) maps of cells with diffraction-limited resolution. xSCYTE is built on a versatile quantitative phase microscope, whose label-free imaging nature allows in-situ, long-time, reusable live-cell analysis. Further, with angle-multiplexing illumination and a pre-trained physics-incorporating Deep Neural Network, xSCYTE can map the 3D RI distribution of a cell by acquiring only a single frame that allows rapid image reconstruction and holds the potential for real-time analysis. By flowing large quantities of cells through a microfluidic channel and equipping xSCYTE with a high-speed camera, we have demonstrated an unprecedented 3D imaging throughput of over 20,000 cells/second while providing sufficient morphological information to distinguish different cellular species. The biomedical application potential of xSCYTE has been evaluated by visualizing and quantifying shear-induced 3D transient deformation of red blood cells that can be correlated with several blood pathologies. With these high-speed and high-precision imaging capabilities empowered by artificial intelligence, we envision xSCYTE may open up many new avenues of biomedical investigations and industries, such as multi-omic assays and quality control during cellular therapeutic manufacturing.

## Introduction

Many emerging biomedical and clinical developments, e.g., drug screening, stem-cell therapies, and regenerative medicine[1–4], involve high-throughput cell analysis[5]. Current solutions are mainly based on flow cytometers, which provide only low-dimensional cellular information hindering more comprehensive cell analysis related to morphology and biophysical properties[6–8]. In recent years, image cytometers[9], including ImageStream[10], have been developed and successfully utilized for morphological and multiparametric analysis of stem cells and senescent cells[8]. Nonetheless, most current image cytometers provide two-dimensional (2D) measurements[5,12–15], which cannot fully reveal intracellular structures and biophysical properties as thoroughly and accurately as their three-dimensional (3D) counterparts[16–18] (e.g., confocal microscopy[12,17,19–22], light-sheet microscopy[12,23–27], structured illumination microscopy[18,28,29]and Optical diffraction tomography (ODT)[30–33], refer to Table 1 for a technical comparison. Meanwhile, 3D imaging methods usually require long scanning time with complex hardware and extensive data processing, while their throughput is typically limited to <1,000 cells/second[18,23], including those based on faster raster scanning hardware[34,35], sample rotation using shear flow[13] utilizing light field microscopy[13,36,37], or more effective reconstruction algorithms[38–40]. Furthermore, many emerging clinical scenarios, such as stem-cell therapy investigations[41] and CAR-T cell screening[42] require in-situ cell analysing and cells may need to be preserved for repeated characterizations. Although fluorescence labelling is commonly used in cytometry, it has several drawbacks[43–46]: (i) photodamage and photobleach caused by fluorescence labelling often prevent long-term imaging; (ii) additional reagents and cell preparation steps are required that make imaging more costly and inefficient; (iii) chemical labels may confound the interpretation, alter cellular structures, and jeopardize cell viability[43,44]. Several promising label-free imaging-based cytometry techniques have been developed for biophysical phenotyping of cellular states[47–50], such as detecting infectious diseases by scrutinizing population composition changes of leukocytes[51], investigating the etiology of malaria[52] and sickle cell disease[48] by mapping membrane fluctuations or deformations of Red Blood Cells (RBCs). However, developing label-free 3D image cytometers with high-throughput faces many challenges. Thanks to latest developments in machine learning, particularly deep learning, it has offered the potential to significantly enhance imaging performance and accelerate data processing. Much progress has been made in applying deep learning techniques to increase imaging throughput and performance[53–55], as demonstrated in super-resolution optical imaging[56], low-photon imaging[57], label-free imaging with computational specificity[53,58–60], etc. Hence, pairing advanced machine learning algorithms with innovative imaging hardware designs has the potential to create more efficient 3D imaging strategies and instruments.

Here, we report eXpress Single-frame CYtometer through Tomographic phasE (xSCYTE) for imaging unlabelled cells in 3D with microsecond-level temporal resolution. Instead of acquiring a large number of images, usually over 40, at different illumination angles or sample depth-scanning positions[61,62], only a single interferogram that multiplexes four illumination angles is captured in xSCYTE. A pre-trained physics-incorporating Deep Neural Network (DNN) is adopted to rapidly map the 3D refractive-index (RI) distributions of cells quantitatively with diffraction-limited spatial resolution from a single interferogram.

Concurrently, 3D image reconstruction time is improved by over 100 times compared with conventional model-based algorithms. With these innovations, we have demonstrated that xSCYTE can achieve a 3D imaging speed of 12,500 volumes/second (vps) when equipped with a high-speed camera. By fast-flowing cells in customized microfluidic devices, xSCYTE is capable of imaging over 20,000 cells/second. By segmenting the cells and extracting a set of 3D morphological and biophysical parameters (i.e., volume, surface area, dry mass, and mean RI), we have shown that xSCYTE can distinguish different cell species during large scale cell characterization. Furthermore, we applied xSCYTE for characterizing transient 3D deformations of RBCs induced by the shearing force in a microfluidic channel on the sub-millisecond scale, which showcases its potential for quantifying cell mechanical properties and monitoring ultra-fast cellular dynamics. By fully unleashing its potential for efficient and high-speed 3D analysis of unlabelled cells, we envision xSCYTE may promote many emerging biomedical investigations and related industries, and subsequently contribute to the development of novel medical diagnostic and treatment techniques in the future.

## Results

### Overview of xSCYTE

The overall pipeline of xSCYTE, as illustrated in Fig. 1a, contains two key components: (i) a quantitative phase microscope with angle-multiplexing optics that simultaneously illuminates the samples from four angles; (ii) a Machine-Learning (ML) engine that converts a single interferogram containing four illumination angles of the cells to a 3D RI map, while compensating for missing spatial frequency information due to the use of only a few illumination angles. Since xSCYTE acquires volumetric information from a single 2D interferogram, the volumetric imaging rate is only limited by the camera frame rate and the number of photons received. With a high-speed camera and adequate illumination power that is still safe for the cells, one can easily push the 3D acquisition speed to over 10,000 vps.

In xSCYTE, we implement an off-axis digital holography design for the quantitative phase microscope[31,63]. Multiplexing four illumination angles for single-frame 3D imaging is achieved by overlapping multiple Lee hologram patterns[64] on a Digital Micro-mirror Device (DMD). The details of the imaging system design are elaborated in *Methods*. In the following sections, we will show that the utilisation of four angles strikes a good balance between acquisition time and reconstruction fidelity.

The ML engine reconstructs the 3D RI map of the sample as follows. First, the raw multiplex interferogram is pre-processed through a spectral filtering method to extract four Phase Approximants (see *Methods*and *Supplementary Material*, Section 1), which are essentially approximations of the quantitative phase delay accrued after the light has gone through the sample at each corresponding angle. The Phase Approximants are then input to a pre-trained DNN model (Fig. 1b (iii)) to infer the 3D RI maps. Thereafter, a linear fitting procedure is applied to recover the quantitative RI values (see *Methods*).

Central to getting xSCYTE to work is the supervised training scheme for the ML engine. We first construct a dataset for training and testing our DNN model (Fig. 1b (i) and (ii)). For each cell, two types of raw data are acquired with our quantitative phase microscope-based imaging platform: (i) 49 sequential interferograms, each from a single scanning illumination angle; and (ii) a single-frame interferogram from illuminating the sample simultaneously with four beams with the same elevation angle and azimuthal angles at $0^{\circ}$, $90^{\circ}$, $180^{\circ}$, and $270^{\circ}$ (refer to *Supplementary Material*, Section 1 for details on the angle scanning patterns in each data type). Acquisition (i) is used to obtain the ground truth 3D RI maps of the cells for training, whereas acquisition (ii) is the normal operating mode of xSCYTE as described earlier.

The ground truth 3D RI maps are produced from acquisition (i) as follows: firstly, phase maps corresponding to each illumination angle are retrieved based on the Fourier transform method. The 49 phase maps are then used to reconstruct 3D RI maps using the Learning Tomography Beam Propagation Method (LT-BPM)[65,66] (see *Supplementary Material*, Section 2). The reliability of LT-BPM as ground truth is validated on calibration samples, including polystyrene beads and 3D-printed cell phantoms (see *Supplementary Material*, Sections 3). The DNN model is based on the Learning to Synthesize by DNN (LS-DNN)[67] principle, which here we generalize for 3D RI reconstruction (see *Methods*). The supervised training procedure minimizes the Negative Pearson Correlation Coefficient (NPCC) loss[68] between the ground truth RI map and the output of the LS-DNN when its inputs are the Phase Approximants. The choice of NPCC is meant to further ensure the preservation of the sample's fine features. The key contribution of the LS-DNN scheme to our problem is to combat the uneven fidelity of low and high spatial frequencies that often occur in DNN training[69]. Thus, the spatial resolution of the final 3D RI maps can be greatly improved.

The DNN is trained with 900 pairs of ground truth RI maps and corresponding Phase Approximants extracted from raw multiplex interferograms of NIH/3T3 cells. The NIH/3T3 cells used for training are cultured on glass-bottom well plates during the data acquisition. Interferograms corresponding to 49 illumination angles are exploited for reconstruction (acquired at 5000 frames/second). The cells can be considered static during one complete angle-scanning process. The ML-engine's ability to generalize beyond its training to different cell types is discussed further in Section 2.3. Moreover, after training, xSCYTE's ML engine is fast in inferring the 3D RI map from Phase Approximants, taking only 0.68 second/volume on average (refer to Supplementary, Section 5 for detailed analysis of computational time and comparison with exiting 3D RI reconstruction methods).

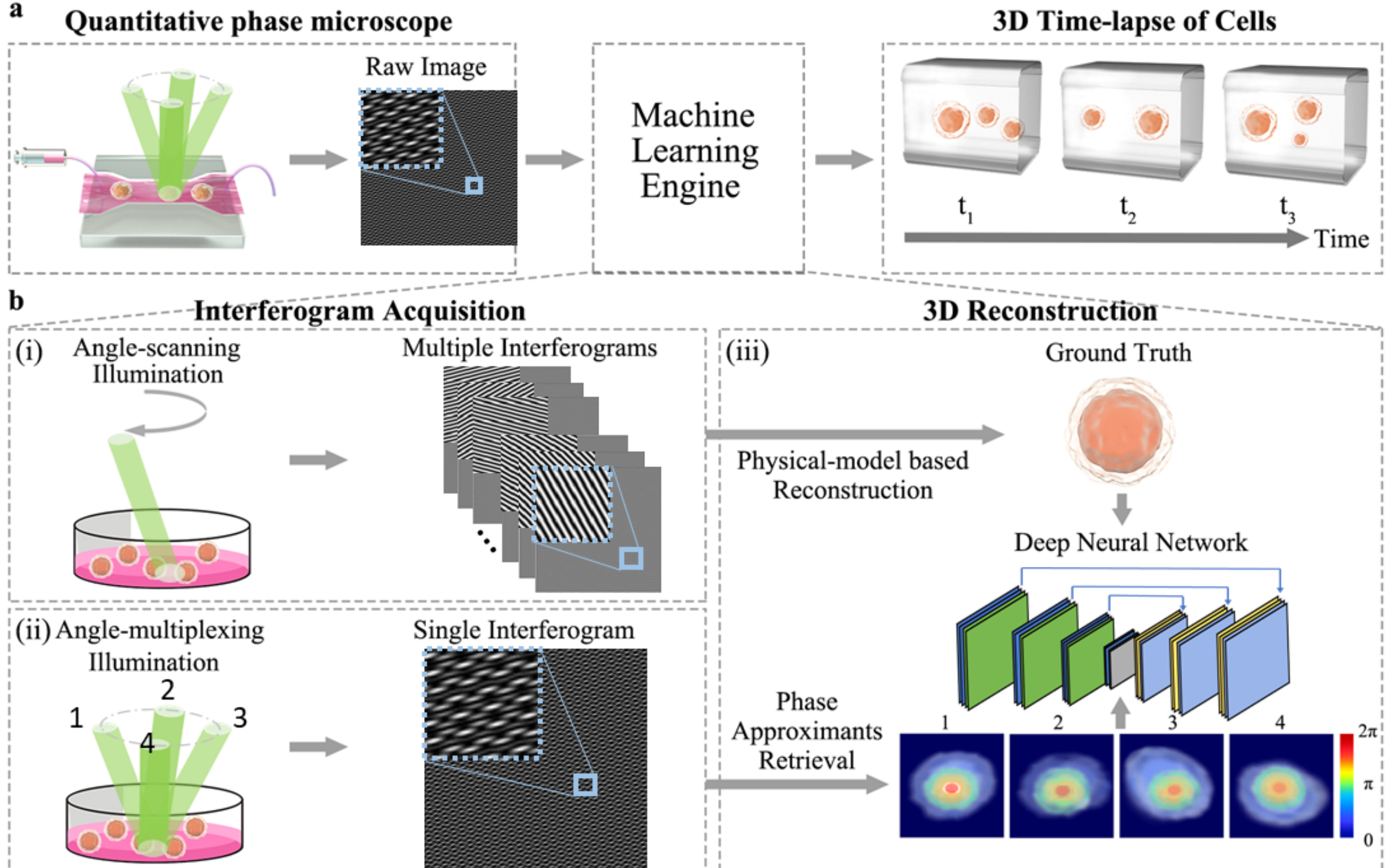

Figure 1. The overall working principle of xSCYTE. (a) The pipeline of 3D time-lapse imaging via xSCYTE consists of two principal steps: multiplex interferograms captured from the quantitative phase microscope and 3D RI map inference from a machine learning engine containing a Phase Approximant retrieval algorithm and a pre-trained DNN. (b) The training process of the machine learning engine for 3D reconstruction: (i) illustrates the acquisition of all 49 interferograms captured under the angle-scanning illumination scheme used for generating ground truth RI maps; (ii) shows the acquisition of the multiplex interferogram; and (iii) describes how the ground truth 3D RI maps (prepared using a physical-model based reconstruction algorithm, i.e., LT-BPM) and the four Phase Approximants (estimated from the multiplex interferogram) are used for training the DNN.

## Quantitative validation

We now turn to the validation of our design choices, namely: (i) the extent we can compress the number of angle-scanning measurements with deep learning; and (ii) the feasibility of multiplexing four illumination angles in one interferogram acquisition. The performance of LT-BPM (i.e., the physical model for acquiring the ground truth RI maps) with sequentially scanning illumination angles drops dramatically when the image acquisition number decreases from 32 to 1 as shown in Fig. 2a-c, where Pearson Correlation Coefficient (PCC), Mean Absolute Error (MAE), and Root Mean Square Error (RMSE) are used for quantitative assessments by comparing with 49-angle LT-BPM. The loss of fidelity can be mitigated by applying deep learning. By pre-training a DNN using ground truth RI maps from LT-BPM with 49 scanning angles, we can predict the 3D RI maps of cells with compressed measurements. We call this method Multi-frame Deep-learning ODT (MDODT) (see *Supplementary Material,* Section 4). Even with only four scanning angles or four acquisitions, MDODT reconstruction results are almost indistinguishable from those of 32 image acquisitions with LT-BPM. However, reducing the number of illumination angles to one significantly deteriorates the performance.

Moving forward, we multiplex four illumination angles into one interferogram using a DMD in an angle-scanning quantitative phase microscope. With only one acquisition, it is found that the imaging performance is similar to MDODT when the number of imaging acquisitions is four, as shown in Fig. 2a-c. Therefore, our choice of multiplexing four illumination angles in xSCYTE is well justified. Note that multiplexing more than four illumination angles in xSCYTE may achieve similar performance, but this will require a new set of training data and re-train the DNN.

**xSCYTE generalizability: 3D RI maps of diverse cell species**

3D RI maps of various cell species are reconstructed with xSCYTE and shown in Fig. 2d-e. The y-z sections at the origin of the x-axis and x-y cross-sections at the centre layer and layers located ±1 μm above and below are shown in different columns in Fig. 2d. 3D renderings of RI maps are also provided in Supplementary Videos 1-4. In Fig. 2e, the nucleoli, the nuclei's boundaries, and other organelles are readily distinguishable in the 3D renderings. Comparisons between xSCYTE and corresponding ground truth reconstructions of various cell species demonstrate the accuracy of xSCYTE (refer to *Supplementary Material,* section 6).

By applying this algorithm to other cell types and evaluating xSCYTE's reconstruction performance on different cell species, we can assess the generalizability of our approach. From the testing results measured with PCC, MAE, or RMSE, it is found that the generalization performance on several other similar eukaryotic cell species (HEK293T, HeLa, COS-7 cells) is comparable to the testing results of NIH/3T3 cells. The same applies when xSCYTE is tested on RBCs, a very different cell type without nuclei and organelles. Although, unsurprisingly, RBC reconstruction is slightly less accurate when reconstructed using NIH/3T3 trained DNN, it mostly maintains the accuracy of cell parameter extraction (eccentricity, volume, etc), as shown in *Supplementary Material,* section 9. The ability to generalize to other types of cells is remarkable, especially given that our ML engine is trained on only ~900 NIH/3T3 cells.

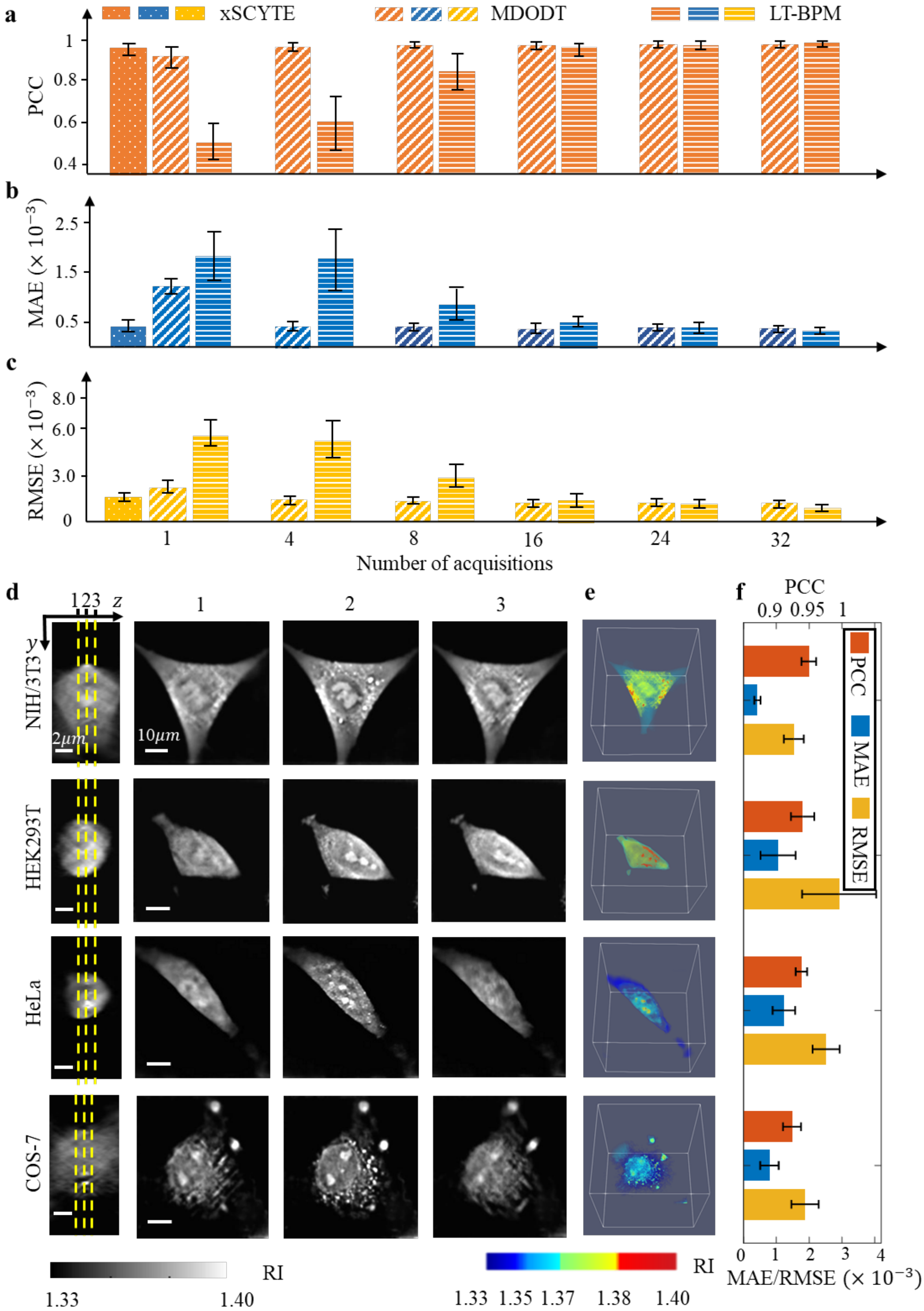

Figure 2. Evaluation of xSCYTE 's performance. Comparison of quantitative metrics (a) PCC, (b) MAE, and (c) RMSE for xSCYTE, MDODT, and LT-BPM as a function of the number of acquisitions. (d)-(f), visualization and quantitative evaluation of predicted RI maps of NIH/3T3 cells, HEK293T cells, HeLa cells, and COS-7 cells by xSCYTE. The first column of (d) shows the cross-sections of the 3D RI map on $y-z$ plane; The $x-y$ cross-sections at different z locations (1) $z_0 - 1\mu m$, (2) $z_0$ and (3) $z_0 + 1\mu m$ (identified with yellow dashed lines in the first column, where $z_0$ indicates the central plane) are shown in the second to fourth columns of (d), respectively. 3D renderings of the RI maps are shown in (e). The same quantitative metrics as in (a)-(c) are shown in (f). 3D rendering videos are provided in Supplementary Videos 1-4.

**High-throughput and high-content 3D cell imaging**

For high-throughput and high-content 3D imaging of cells, we incorporate a high-speed camera and inject cells in microfluidic channels (see *Supplementary Material*, Section 10, and Fig. S9 c-e for the cross-sections of the channels). The speed of imaging that yields full 3D RI maps is 12,500 vps with only 10-15 μs exposure time. Three types of cell specimens are used with the following suspension and flow velocity conditions: NIH/3T3 cells with suspension at $20\times10^6$ cells/mL and flow velocity at 0.46 m/sec, Jurkat T cells with suspension at $90\times10^6$ cells/mL and flow velocity at 0.31 m/sec, and mouse RBCs with suspension at $100\times10^6$ cells/mL and flow velocity at 0.56 m/sec. 3D renderings of RI maps at different time points for flowing NIH/3T3 cells, Jurkat T cells, and mouse RBCs are shown in Fig. 3a-c, respectively. A total of 936 NIH/3T3 cells are acquired within 160 ms (equivalent to a throughput of approximately 5,850 cells/second) and rendered into a time-lapse video (Supplementary Video 5). The Jurkat T cells and mouse RBCs are prepared at a higher cell density to push the throughput limit. A total of 3,350 Jurkat T cells are captured within 160 ms (refer to the rendered time-lapse video in Supplementary Video 6), while a total of 1,400 RBCs are captured within 80 ms (refer to the rendered time-lapse video in Supplementary Video 7). Therefore, we have reached a throughput of approximately 20,938 cells/second for Jurkat T cells and a throughput of 17,500 cells/second for mouse RBCs. Note that our demonstrated cell measurement throughput is 100-1000 times faster than other current 3D cell imaging methods.

By segmenting the cells from their 3D RI maps, we can extract a set of morphological and biophysical parameters, as well as the dry mass that reflects the total cell protein content. The distributions of these quantities in the entire cell population could be subsequently used for cell characterization and classification. As a proof-of-concept study, we extracted the mean RI, volume, surface area, and dry mass of the captured NIH/3T3 cells, Jurkat T cells, and mouse RBCs and explored their distributions for differentiating the cell populations. For each cell type, 500 cells are randomly chosen from the corresponding cell population as captured in the 3D time-lapse videos. From volume & surface area distribution (Fig. 3d), as expected the cell sizes from small to large are mouse RBCs, Jurkat T cells, and NIH/3T3 cells. We also noticed that mouse RBCs and Jurkat T cells have narrower variances for these measured parameters than NIH/3T3 cells do. This difference is also expected as NIH/3T3 cells may be harvested from different cell cycle stages, while RBCs and Jurkat T cells are mature cells possessing more uniformly distributed cell sizes. In Fig. 3e, we plot dry mass & volume distribution, where the three cell types are well separated. The linear relationships of dry mass *vs.* volume indicate that NIH/3T3 cells and mouse RBCs have similar RI mean values, while Jurkat T cells' RI values are slightly lower. To further characterize the cells using the 3D shape information, the correlations between mean RI and volume-to-area ratio, as well as dry mass and mean RI, are explored as presented in Fig. 3f-g, where a clear separation of all the three cell types is observed. The details for computing mean RI, volume, surface area, and dry mass and their distributions are elaborated in *Supplementary Material*, Section 8.

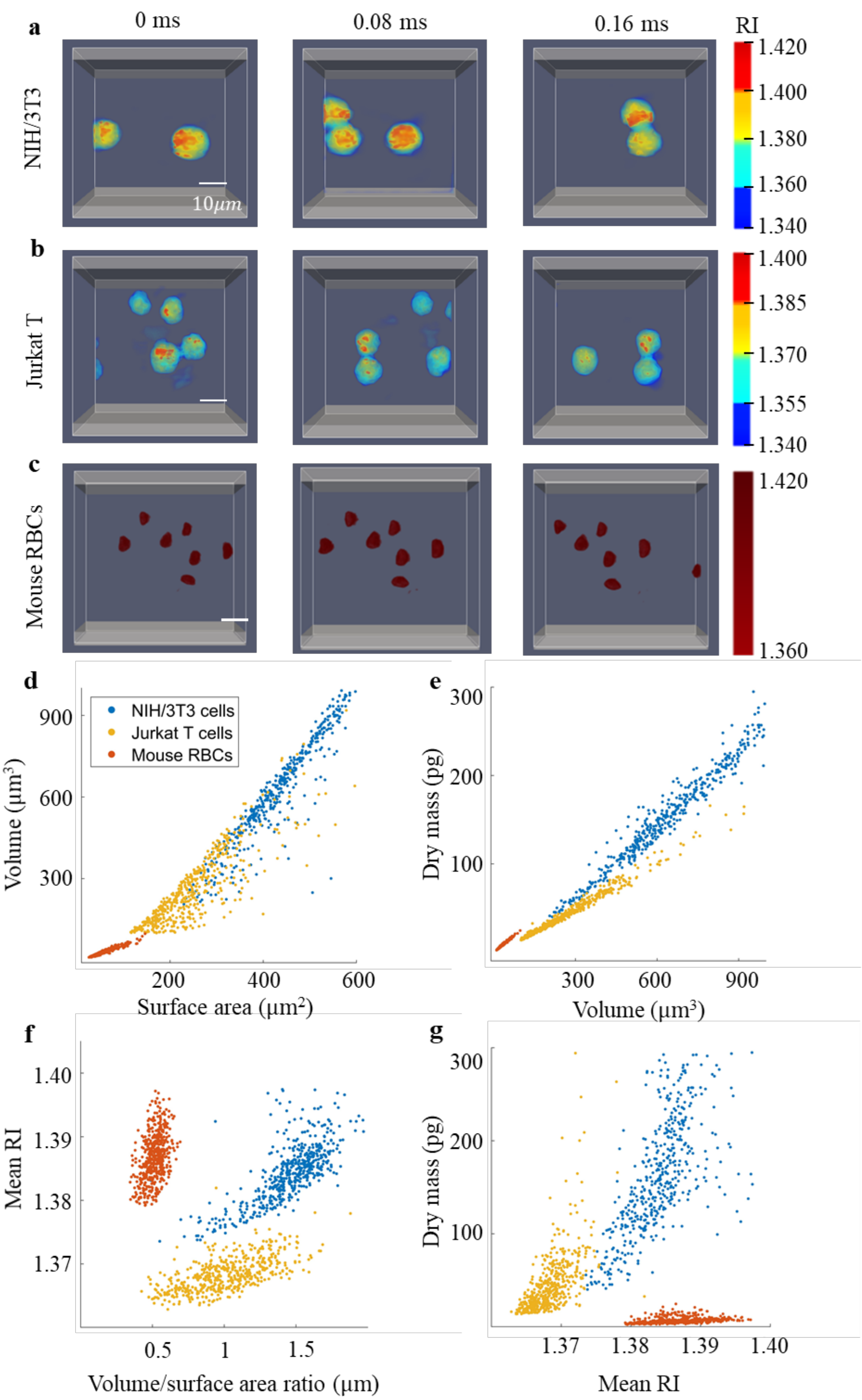

Figure 3. Demonstration of xSCYTE as a high-throughput and high-content 3D imaging flow cytometer. 3D rendering of the imaged (a) NIH/3T3 cells, (b) Jurkat T cells, and (c) mouse RBCs at different time points as they flow through the channels, respectively. The flow velocities of the NIH/3T3 cells, Jurkat T cells, and mouse RBCs are 0.46 m/sec, 0.31 m/sec, and 0.33 m/sec, respectively. The imaging speed is 12.5k vps. (d)-(g), the scatter plots of (d) surface area vs. volume, (e) volume vs. dry mass, (f) volume/surface area ratio vs. mean RI, and (g) mean RI vs. dry mass. The data from NIH/3T3 cells, Jurkat T cells, and mouse RBCs are shown with blue dots, orange dots, and red dots in (d)-(g).

## Observing red blood cell 3D deformation

As illustrated in the scheme of shear force-induced RBC deformation in Fig S9f, the RBCs in the microfluidic channel (see *Supplementary Material*, Sections 10, and 11) can be deformed by increasing shear rates when they float from wider to narrower regions (a process similar to RBCs travelling in capillaries in human). The ability to observe RBC mechanical deformation is important as their biomechanical properties have been linked to several blood pathologies[47,48,51,52].In our RBC experiment, the flow velocity is set at 0.33 m/sec, and the 3D image acquisition speed is 10,000 vps. The entire process of RBC deformation is captured by xSCYTE for a total of 5 ms, and a time-lapse video is created to visualize the whole process (see Supplementary Video 8). Volumetric renderings of selected frames are shown in Fig. 4a. The ML engine used for the results in this section was trained with a dataset of ∼500 input-ground truth pairs of human RBCs with PCC approaching ∼0.96 on test samples. We also trained the ML engine with NIH/3T3 cells and compared the RBC results with the current results (refer to the details in *Supplementary Material*, Section 9). The RBC results trained with NIH/3T3 are slightly worse with PCC drops to ~ 0.88, while the extracted morphological parameters remain similar, which further supports the generalization capability of xSCYTE.

Figure 4b quantifies the evolution of a selected RBC's eccentricity (formula in *Supplementary Material*, Section 8) as it drifts into the region of higher shear rate in the microfluidic channel. As expected, the RBC gets elongated as it flows through the transition region, and it stabilizes after completely entering the narrower section of the channel.

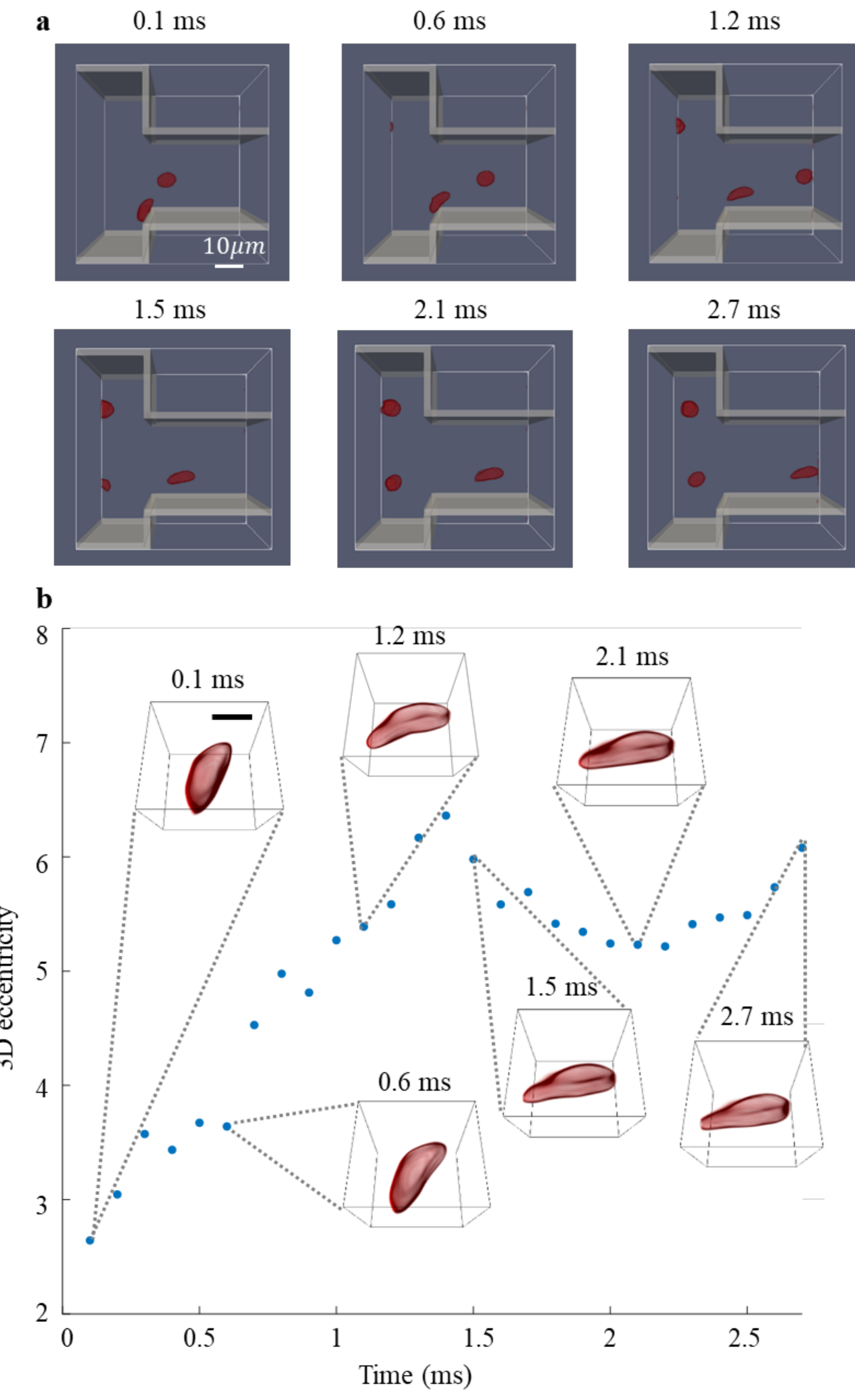

Figure 4. Demonstration of xSCYTE for 3D visualization of RBC deformation in a microfluidic channel. (a) 3D renderings of the RBCs flowing in the microfluidic channel at time points of 0.1, 0.6, 1.2, 1.5, 2.1, and 2.7 ms. (b) Variation of the 3D eccentricity of a selected RBC over time (0-2.7 ms). 3D renderings of the selected RBC at time points 0.1, 0.6, 1.2, 1.5, 2.1, and 2.7 ms are embedded into this figure. The complete flow process for a total of 5 ms is provided in Supplementary Video 8.

## Discussion

xSCYTE's performance is based on careful system-level coordination of our two key design choices: multiplex illumination with four angles, a relatively small number, to obtain the raw images; and a physics-incorporating ML engine that converts the raw images to Phase Approximants and finally to 3D RI maps. The tests described in the previous sections and in the *Supplementary Material* are meant to validate 3D RI map fidelity despite operating the instrument aggressively at tens of thousands of volumes per second. Concerns about ML, in particular, are often well-justified when the parameters of an algorithm are determined not from first principles directly but rather from numerical optimization—this is the infamous "black box" problem. Traditional compressed sensing algorithms[70–72] often come under similar criticism. In our case, by incorporating the Phase Approximant as a prior from instrument physics and the Learning-to-Synthesize scheme for spatial frequency rebalancing, we sought to mitigate the black box concern explicitly. Besides, ML also enhances the spatial resolution of xSCYTE for better resolving the intracellular structures and more accurately extracting morphological parameters.

By further exploiting and extending the capabilities of xSCYTE, it may support many new biomedical investigations and cell-based industrial applications in the near future. First, 3D RI map reconstruction by xSCYTE can provide rich information on both the morphology and biochemical content of the cells, facilitating the development of more efficient ML-based image classification algorithms for cell discrimination, which has applications in multi-omic assays that require sampling of a large cell population. Second, more powerful generative AI techniques may empower xSCYTE with higher imaging reconstruction performance[6] and more effective cell classification/segmentation[73], while transfer learning[74] can be used to further broaden the diversity of cells that we can reconstruct reliably. These improvements may lead to more critical applications, such as circulating tumour cell detection[75], leukocyte sub-type differential counting[46], drug screening[76], etc. Third, by incorporating cell sorting in xSCYTE[6,15], we may create a powerful tool to tackle several critical biomedical applications, such as regenerative medicine by providing quality control of cells during cellular therapeutic manufacturing and in vitro fertilisation (IVF) by offering 3D motion and viability analysis of sperms. However, the tremendous data flow is a major obstacle for bringing these ideas into reality. As more powerful computational hardware, such as graphic processing units (GPUs), high bandwidth memory, and high density storage are continuously produced, they give us the confidence that the data processing needs in 3D image cytometry can be overcome, not to mention the recent advancements in optical computing that have shown considerable promise in overcome the bandwidth limit of electronic systems[77–80].

# Methods

Here, we elaborate on the basic principles of the xSCYTE experimental platform and algorithmic pipeline introduced briefly in section 2.1. Section 4.1 describes the quantitative phase microscope apparatus that acquires multiplex interferograms with a high-speed camera. Section 4.2 discusses the estimation of Phase Approximants from the multiplex interferogram. The design of LS-DNN and training strategy for xSCYTE's ML engine are provided in Section 4.3. The technical descriptions of Sections 4.1-3 are also graphically summarized in Fig. 5. Brief technical introductions to alternative ODT methods *vis-à-vis* xSCYTE are in section 4.4. Cell preparation protocols are described in Section 4.5. Additional technical details and analysis of experimental results are in *Supplementary Material.*

**High-speed angle-scanning and angle-multiplex quantitative phase microscope**

The schematic of the quantitative phase microscope apparatus used in xSCYTE is shown in Fig. 5a. A 532 nm laser (CNI Lasers, MGL-III-532-300mW) is used as the illumination source. The laser beam is divided into two beams by a 1×2 Single-Mode Fibre Coupler (SMFC). One beam serves as the reference for interferometric detection, while the other is directed to the sample. The sample beam is collimated by a lens L1 ($f_1$ = 200 mm) before impinging onto DMD D1 (Texas Instruments Inc., DLP LightCrafter 9000), which is programmed for displaying Lee hologram patterns consisting of multiple diffracted plane waves. Lens L2 ($f_2$ = 150 mm) enables these reflected beams to form a series of diffraction spots at the Fourier plane, where the second DMD2 (Texas Instruments Inc., DLP LightCrafter 6500) is placed. The filter mask patterns shown in inset 1 of the figure are loaded onto DMD2 to block spurious diffraction orders and only allow downstream the desired 1$^{st}$ diffraction order. Next, the filtered beam is collimated by lens L3 ($f_3$ = 200 mm), followed by a 4f system composed of a tube lens L4 ($f_4$ = 300 mm) and an objective lens OL1 (Zeiss, 63X/1.3, water immersion). The 4*f* system magnifies the angular range of the sample beam. After incidence on the sample, the scattered light is collected by the objective lens OL2 (Zeiss, 63X/1.25, oil immersion), then reflected by mirror M1 and collimated by lens L5 ($f_5$ = 150 mm). A Beam Splitter (BS) behind lens L5 combines the sample and the reference beams into the multiplex interferogram, which is spatially magnified by the 4*f* system consisting of lens L6 ($f_6$ = 60 mm) and L7 ($f_7$ = 400 mm). The resulting raw image is captured by a high-speed camera (Photron, Fastcam SA-X2).

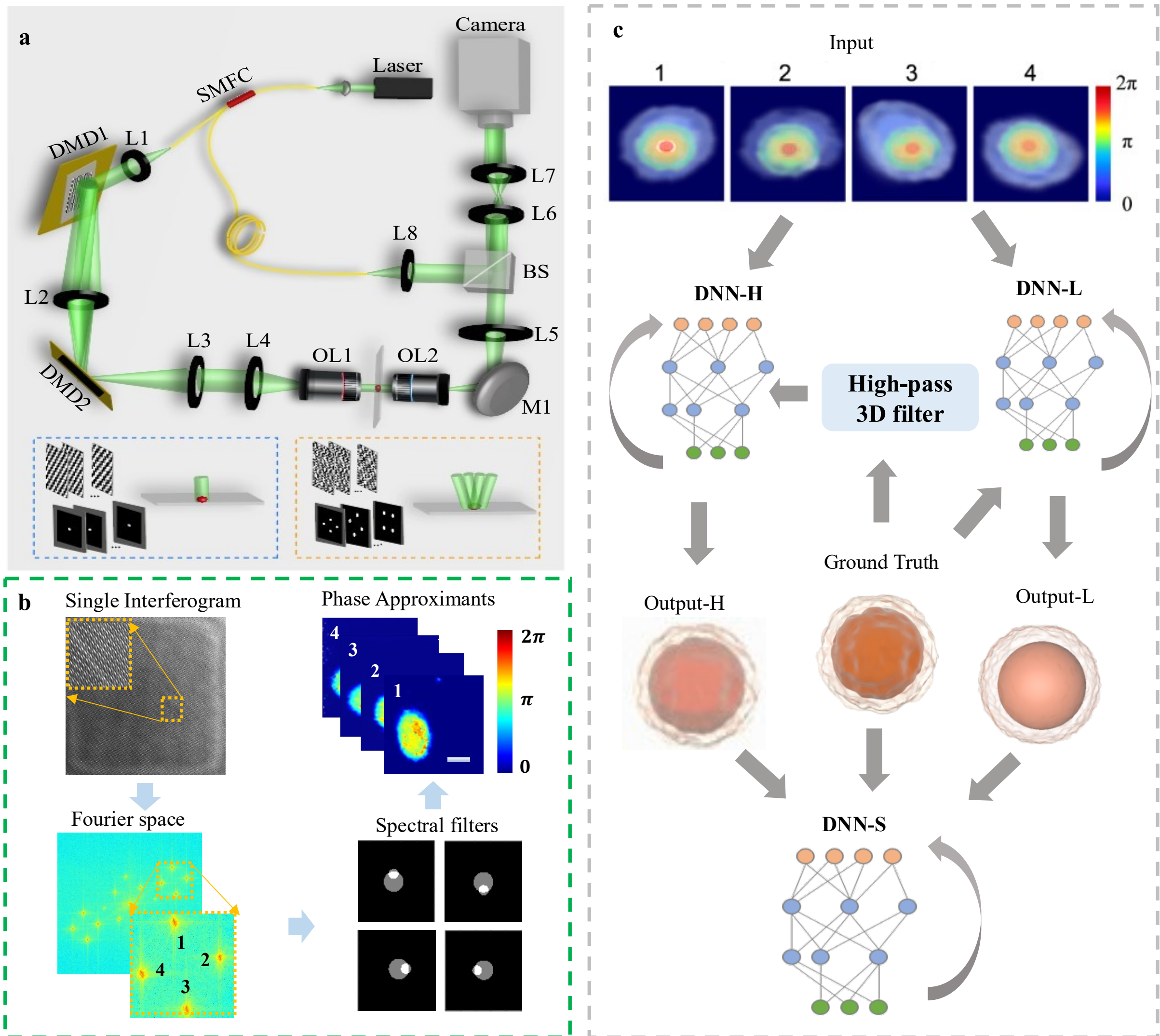

Figure 5. Detailed depiction of the operation of xSCYTE. (a) System design of xSCYTE and illustration of single illumination and multiplex illumination. (b) Pipeline of the spectral filtering method for acquiring the Phase Approximants. (c) Flow chart of the two-step training process for LS-DNN.

### Phase Approximant retrieval

The multiplex interferogram is first spatially Fourier transformed as illustrated in Fig. 5b. Nine bright spots can be seen near the central region of the Fourier space. The central spot is the $0^{th}$ diffraction order or Direct Current (DC) order, while the other eight spots are pairwise cross-correlations of the four scattered beams and the reference beam. The top right and bottom left regions are the $+1^{st}$ and $-1^{st}$ orders which correspond to interference between each one of the four scattered beams and the reference beam. To extract the phase maps of the four illumination angles individually, four specially designed spectral filters are digitally applied to the $+1^{st}$ order. The retrieved Phase Approximants are shown in Fig. 5b. The mathematical formulation of the interferogram and the retrieval of the Phase Approximants are in the *Supplementary Material,* Section 1.

### Learning to Synthesize by DNN

Our motivation to develop this technique is based on earlier observations that 2D reconstructions by machine learning often exhibit deficiencies at the high end of the spatial frequency spectrum. This has been attributed to the relative sparsity of high frequencies in training databases[69]. To compensate, the two-step LS-DNN algorithm splits the spectral information into two bands, high and low, processes them separately, and then recombines. This requires three DNNs, which are trained separately, as shown in Fig. 5c: one trained for the high-frequency bands, DNN-H; one trained for the low-frequency bands, DNN-L; and a final "synthesizer" DNN-S trained to output the compensated reconstruction exhibiting even fidelity at all frequency bands[67].

Here, we modified the previously 2D-oriented LS principle[67] to work for 3D RI reconstruction from Phase Approximants, as follows. Let $n(x,y,z)$ denote the RI map in the Cartesian coordinates $(x,y,z)$ with $z=0$ corresponding to the centre of the reconstructed 3D RI map. In xSCYTE, the ground truth 3D RI maps contain 100 layers along the *z*-direction, and the interval between adjacent layers is 0.21 $\mu m$. As the cells that we use are mostly confined to a small axial dimension within $|z| \leq 8\ \mu m$ (equivalently to 80 layers), and in the cell support most sub-cellular features are located within $|z| \leq 6\ \mu m$ (i.e., the region where high frequencies are of the highest significance). In the region $6\ \mu m < |z| \leq 8\ \mu m$, relatively fewer fine details of the cells are present, whereas in the region $|z| > 8\ \mu m$ is typically void. Accordingly, we define the spatial filter as:

$$M(k_x, k_y; z) = \begin{cases} 1, & |z| > 8\mu m \text{ (no filtering)} \\ (k_x^2 + k_y^2)^{0.8}, & 6\mu m < |z| \leq 8\mu m \text{ (moderate filtering)} \\ (k_x^2 + k_y^2)^{1.5}, & |z| \leq 6\mu m \text{ (strong filtering)} \end{cases} \quad (1)$$

Let $F(\cdot)$denote the Fourier transform operator. We produce filtered RI functions $\tilde{n}(x,y,z) = F^{-1}(F(n(x,y,z))M(k_x,k_y;z))$ for use in the DNN-H pipeline, both training and in actual operation. For the DNN-L pipeline, we use the unfiltered $n(x,y,z)$. The DNNs are trained separately in the supervised mode, as in the previous work of 2D LS-DNN[67].

The aforementioned LS-DNN architecture works well for the training of NIH/3T3 cells as we discussed earlier in Section 2. Although the ML engine trained with only NIH/3T3 cells has satisfactory generalization ability, training on other species of cells is sometimes necessary to further improve the performance of xSCYTE. For example, we can fine tune the hyperparameters in the NIH/3T3 cell trained DNN model by training on a small number of other cell species (e.g., transfer learning). Since different types of cells have various cellular structures, our training strategies on different species of cells are distinct. For example, RBCs have simpler structures and smaller sizes than NIH/3T3 cells. Therefore, we could only use the input-ground truth pairs of RBCs to train DNN-L (see *Supplementary Material,* Section 9). Furthermore, the hyperparameters of the high-pass filters can be changed to accommodate the training of different species of cells.

To design xSCYTE's ML engine, we used a dataset consisting of 900 input ensembles obtained from NIH/3T3 cells. Each ensemble consists of four Phase Approximants estimated from the corresponding multiplex interferogram and the ground truth for the same scene, *i.e.,* the 3D RI map reconstructed from 49 true phase maps with LT-BPM (see *Supplementary Material*, Sections 2&3). From the dataset, 5% of the ensembles are used for validation and a further 39 disjoint ensembles are reserved for testing. The remaining ensembles are used for training. For the results shown in Section 2.5, the same procedure is carried out but with RBCs instead of NIH/3T3 cells.

The training loss function is the Negative Pearson Correlation Coefficient (NPCC), defined as:

$$\mathrm{NPCC}(n,\hat{n}) = -\frac{\sum_i (n_i-\bar{n})(\hat{n}_i-\bar{\hat{n}})}{\sqrt{\sum_i (n_i-\bar{n})^2 \sum_i (\hat{n}_i-\bar{\hat{n}})^2}}, \quad (2)$$

where $n$ and $\hat{n}$ are the ground truth and the output of a neural network, respectively; $\bar{n}$ and $\bar{\hat{n}}$ are their means; and $i$ indexes the voxels. This choice has been previously proven capable of reconstructing fine features with good fidelity[69]. NPCC is invariant under linear transformations, *i.e.,* $\mathrm{NPCC}(n,\hat{n}) = \mathrm{NPCC}(n, a\hat{n}+b)$ for all $a$ and $b$ values. Therefore, to obtain RI distributions at the correct scale, in the validation examples we linearly fit the training ground truth to the neural network output and estimate the coefficients $\alpha_1$ and $\alpha_2$ from the least squares (refer to *Supplementary Material*, Section 7). The estimated values of $\alpha_1$ and $\alpha_2$ are then fixed for the subsequent operation in xSCYTE, and the final quantitative 3D RI map is produced as $\hat{n}_{\mathrm{final}}(x,y,z) = \alpha_1 \hat{n}(x,y,z) + \alpha_2$.

**Summary of 3D image cytometry methods**

Here, we provide a summary of current state-of-the-art developments in 3D optical imaging of cells. Widely used confocal microscopy[12,17,19–22], light-sheet microscopy[12,23–27], and structured illumination microscopy[18,28,29] can offer excellent optical sectioning abilities. However, the cumbersome image scanning apparatus is an obstacle for available high-throughput 3D cell imaging, while most of these methods are based on fluorescence imaging. Optical diffraction tomography (ODT)[30–33] is another promising approach; however, it relies on the rotation of the imaged objects (cells) or scanning the beam or focal depth, which makes the system complex it difficult to realize high-throughput 3D cell imaging. xSCYTE offers an outstanding 3D cell imaging throughput ($> 20{,}000$ events/cells/particles per second) with a spatial resolution of $\sim 0.5\mu m$. The throughput even surpasses the most state-of-the-art 2D image cytometers[15,49,55,81–83]. A detailed technical comparison is summarized in Table 1.

Table 1: Comparison of widely used 3D cell imaging techniques with xSCYTE

| Imaging method | Spatial resolution | Cell measurement throughput |
| --- | --- | --- |
| Confocal[12,17,19–22] | $\sim 0.5\ \mu m$ | ~ 1,000 events/cells/particles per sec |
| Light-sheet[12,23–27] | $\sim 1\mu m$ | ~ 2,000 events/cells/particles per sec |
| Structured illumination[18,28,29] | $\sim 1\mu m$ | ~ 800 events/cells/particles per sec |
| Optical diffraction tomography[30–33] | $\sim 0.5\ \mu m$ | ~ 100 events/cells/particles per sec |
| **xSCYTE** | $\boldsymbol{\sim 0.5\ \mu m}$ | **>20,000 events/cells/particles per sec** |

## Cell preparation

COS-7, HeLa, NIH/3T3, HEK293T, and Jurkat T cell lines are obtained from American Type Culture Collection (ATCC) and tested free of mycoplasma contamination. COS-7, HeLa, NIH/3T3, and HEK293T cells are cultured in a 6-well plate (SPL) and immersed in high-glucose DMEM (Gibco), supplemented with 10% fetal bovine serum (Gibco) and 1% penicillin-streptomycin (Gibco). Jurkat T cells were cultured in RPMI 1640 medium (Gibco), supplemented with 10% fetal bovine serum (Gibco) and 1% penicillin-streptomycin (Gibco). Cells are passaged every 2–3 days and are incubated at 37°C in a humidified atmosphere containing 5% $CO_2$.

For live-cell imaging, cells are plated in 50 mm ibidi µ-dish with ibiTreat (Ibidi) at 2500 cells/cm$^2$ with a 24-hour long growth. Two hours before imaging, the debris and non-attached cells are removed and washed gently with 1x sterile phosphate buffer saline (PBS). Finally, a complete cell growth medium is added for live-cell imaging.

Red blood cell collection: BALB/c mice are maintained by the Laboratory Animal Service Centre, The Chinese University of Hong Kong, Shatin, Hong Kong SAR. All animal procedures are conducted with the approval of the Animal Experimentation Ethics Committee (Ref. No.: 18/233/MIS) of The Chinese University of Hong Kong and the Department of Health, the Government of the HKSAR under the Animals (Control of Experiments) Ordinance, Chapter 340 (18–522 in DH/SHS/8/2/1 Pt.12 and 18–523 in DH/SHS/8/2/1 Pt.12). For the withdrawal of blood, the mouse is first restrained, and approximately 100 µL blood is collected from the saphenous vein by puncturing with a 25 AWG needle (Becton Dickinson) to a heparinized capillary tube. The collected blood is then washed twice with PBS by centrifugation at 500 × g for 5 min and pellets of RBCs are obtained. Finally, the supernatant is discarded, and RBCs are resuspended with PBS and ready for subsequent experiments.

Expired human RBCs (HA RE001F3) are aspirated from a 200 mL unit of packed RBCs using 23 AWG needle in a 1 mL syringe (Becton Dickinson) and washed twice with PBS by centrifugation 500 × g for 5 min to obtain the pellet of human RBCs. The isolated RBCs are then resuspended in PBS for subsequent experiments.

All investigations are conducted with freshly isolated RBCs (within 4 hours from the collection either from mice or human blood unit). All centrifugations to isolate RBCs are conducted at 4 °C using a high-speed refrigerated centrifuge (Neo-fuge 13 R, Heal Force).

**Acknowledgments:**

The authors gratefully acknowledge Mr. Subeen Pang (Mechanical Engineering Department, MIT) for useful discussions on the realisation of LT-BPM; Dr. Marianne M. Lee (School of Life Sciences, The Chinese University of Hong Kong) for assistance with the mouse studies; Dr. Nelson Tang (Department of Chemical Pathology, Faculty of Medicine, The Chinese University of Hong Kong) for providing the human RBCs; Mr. Michał Ziemczonok (Institute of Micromechanics and Photonics, Warsaw University of Technology) for providing the 3D printed cell phantom samples; and Dr. Ulugbek Kamilov (Department of Computer Science and Engineering and Department of Electrical and Systems Engineering, Washington University in St. Louis) for providing the LT-BPM code.

**Funding:**

Croucher Innovation Awards 2019 grant CM/CT/CF/CIA/0688/19ay (R.Z., Y.H.)

Hong Kong Innovation and Technology Fund grant ITS/098/18FP, ITS/178/20FP, and ITS/148/20 (R.Z., Y.H.)

The Chinese University of Hong Kong Research Sustainability of Major RGC Funding Schemes – Strategic Areas (R.Z., Y.H.)

National Institutes of Health (NIH) grant 5-P41-EB015871-27, 5R21NS091982-02, and the Hamamatsu Corporation (B.G., Z.Y., P.T.C.S.)

Singapore–MIT Alliance for Research and Technology (SMART) Center, Critical Analytics for Manufacturing Personalized-Medicine (CAMP) Interdisciplinary Research Group (B.G., P.T.C.S., G.B.)

MathWorks Fellowship (B.G.)

NIH grant R01DA045549 and R21GM140613-02 (Z.Y., P.T.C.S.)

Research Grants Council of the Hong Kong Special Administrative Region, China project CUHK 14203919 and C5011-19GF (Y. P. H., H. R.)

VC Discretionary Fund, the Chinese University of Hong Kong project 8601014 (Y. P. H., H. R.)

Hong Kong Innovation and Technology Fund grant ITS/133/20 (M. K. C., Y. P. H., H. R.)

Intelligence Advanced Research Projects Activity (IARPA) grant FA8650-17-C-9113 (M.D., Z.W., G.B.)

Singapore–MIT Alliance for Research and Technology (SMART) Center, Retinal Analytics via Machine learning aiding Physics (RAMP) Intra-Create Thematic Grant (G.B.)

**Author contributions:**

R.Z., G.B., P.T.C.S., and B.G. conceived the idea of xSCYTE. Y.H. built the optical system, performed the experiments, and collected the data. B.G. implemented the algorithm for reconstructing the ground truth of 3D refractive index maps. M.D. and B.G. designed and trained the machine learning engine for xSCYTE. B.G. and Y.H. implemented the algorithms for retrieving the biophysical parameters and analysed the data. H.R. designed the microfluidic channel and supported the microfluidic-related experiments. Y.W. and Y.H. cultured the cells required for the experiments. Y.H. created the 3D renderings of the cell images and the videos. B.G., Y.H., M.D., and R.Z. wrote the first draft of the manuscript. R.Z., G.B., and P.T.C.S. supervised the research. All authors contributed to commenting, reviewing, and revising the manuscript.

**Conflict of Interest:** A US non-provisional patent (US20230186558A1) and a China invention patent (CN116263411A) have been filed based on this work. R.Z. is a co-founder of Bay Jay Ray Technology limited.

**Data and materials availability:** All data are available in the manuscript and the supplementary materials. A preprint has been posted on arXiv (arXiv.2202.03627).

All the code are included in the following website: https://github.com/xSCYTE/xSCYTE-code.git